\documentclass[prl,aps,amssymb,superscriptaddress,twocolumn, hypertext,longbibliography]{revtex4-2} 
\usepackage{hyperref}
\hypersetup{
    colorlinks,
    linkcolor={black},
    citecolor={blue!80!black},
    urlcolor={blue!80!black}
}
\usepackage{graphicx}
\usepackage{dcolumn}
\usepackage{bm}
\usepackage{soul}
\usepackage[utf8]{inputenc}
\usepackage{amsmath}
\usepackage{multirow}
\usepackage{xcolor}
\usepackage[per-mode=symbol]{siunitx}
\usepackage{braket}
\usepackage{textcomp}
\DeclareSIUnit\gauss{G}
\graphicspath{{/}}

\makeatletter
\renewcommand*{\@fnsymbol}[1]{\ensuremath{\ifcase#1\or \dagger \or *\or \ddagger\or
   \mathsection\or \mathparagraph\or \|\or **\or \dagger\dagger
   \or \ddagger\ddagger \else\@ctrerr\fi}}
\makeatother

\begin{document}


\title{Atomic spin-controlled non-reciprocal Raman amplification of fibre-guided light}

\author{Sebastian Pucher}\thanks{C.L. and S.P. contributed equally to this work.}

\author{Christian Liedl}\thanks{C.L. and S.P. contributed equally to this work.}

\author{Shuwei Jin}

\author{Arno Rauschenbeutel}

\author{Philipp Schneeweiss}
\email{philipp.schneeweiss@hu-berlin.de}

\affiliation{
 Department of Physics, Humboldt-Universit\"at zu Berlin, 10099 Berlin, Germany
}

\date{\today}

\begin{abstract}
In a non-reciprocal optical amplifier, gain depends on whether the light propagates forwards or backwards through the device.
Typically, one requires either the magneto-optical effect, a temporal modulation, or an optical nonlinearity to break reciprocity.
By contrast, here, we demonstrate non-reciprocal amplification of fibre-guided light using Raman gain provided by spin-polarized atoms that are coupled to the nanofibre waist of a tapered fibre section. The non-reciprocal response originates from the propagation direction-dependent local polarization of the nanofibre-guided mode in conjunction with polarization-dependent atom-light coupling. We show that this novel mechanism does not require an external magnetic field and that it allows us to fully control the direction of amplification via the atomic spin state. Our results may simplify the construction of complex optical networks. Moreover, suitable solid-state based quantum emitters provided, our scheme could be readily implemented in photonic integrated circuits.
\end{abstract}

\maketitle


Non-reciprocal optical devices are paramount in optical technologies. They treat light differently when propagating forwards or backwards~\cite{potton2004reciprocity,jalas2013what,asadchy2020tutorial}, thereby enabling, e.g., diodes and circulators for light. The established ways to break reciprocity are based on the magneto-optical effect~\cite{caloz2018electromagnetic}, temporal modulation~\cite{sounas2017non, fan2018nonreciprocal}, or optical nonlinearities~\cite{asadchy2020tutorial,khurgin2020non,kittlaus2020reply}. Recently, the internal spin of quantum emitters that are coupled to spin-momentum locked nanophotonic modes has also been used to achieve non-reciprocity~\cite{lodahl2017chiral}. This enabled a novel class of non-reciprocal elements that, by now, comprises experimental realizations of isolators~\cite{sayrin2015nanophotonic} and circulators~\cite{scheucher2016quantum}. However, a non-reciprocal amplifier is hitherto missing. For such amplifiers, the optical gain depends on whether the light propagates forward or backward through the gain medium. In terms of applications, non-reciprocal amplifiers for light facilitate the optical read-out of sensitive signal sources such as quantum systems while avoiding undesired optical feedback. Moreover, they simplify the construction of complex optical networks as there is no amplification of undesired reflections. Non-reciprocal gain also naturally lends itself to constructing unidirectional ring lasers.

Examples for recent experimental realizations of non-reciprocal optical amplifiers include schemes based on Doppler shifts in hot atomic vapours~\cite{lin2019nonreciprocal}, on optomechanical effects~\cite{ruesink2016nonreciprocity, fang2017generalized, shen2018reconfigurable}, and on stimulated Brillouin scattering in silicon~\cite{otterstrom2019resonantly}. For technical applications, optical Raman amplifiers are of particular importance. In this context, non-reciprocal amplification using the propagation direction-dependent polarization overlap of Raman light fields guided in nanophotonic waveguides has been demonstrated~\cite{krause2012measurement}. Moreover, non-reciprocal amplification via spin selective photon-phonon interactions enabled by stimulated Raman scattering was theoretically studied recently~\cite{lawrence2019nanoscale}. 
\begin{figure}[tb]
	\includegraphics[width=1\linewidth]{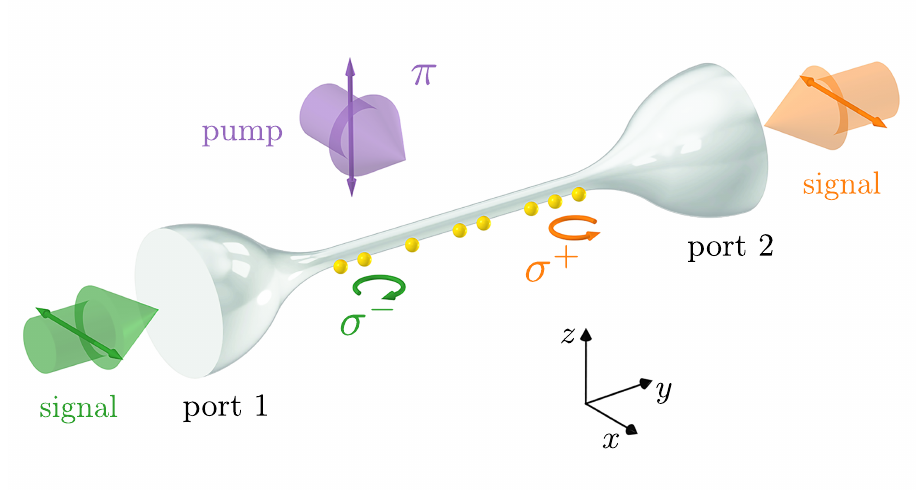}
	\caption{\textbf{Schematic of experimental setup}. Individual caesium atoms (yellow spheres) are distributed over a linear periodic array of optical traps, parallel to the nanofibre waist of a tapered optical fibre. Signal light is launched into this nanophotonic waveguide from port 1 or port 2. We measure the transmitted power and infer the direction-dependent power transmission coefficients $T_{1\rightarrow 2}$  and $T_{2\rightarrow 1}$. Depending on the propagation direction, the evanescent part of the nanofibre-guided signal mode at the location of the atoms is predominantly $\sigma^-$-  or $\sigma^+$-polarized (green and orange circular arrows, respectively), enabling propagation direction-dependent light--matter coupling. The quantization axis is $+z$. A $\pi$-polarized free-space pump laser field (purple) propagates in the $+x$-direction. The signal and pump fields drive two-photon transitions in a Raman configuration. 
}
\label{fig:setup}
\end{figure}

Here, we experimentally demonstrate non-reciprocal amplification of light using laser-trapped, spin-polarized caesium (Cs) atoms, which are coupled to a nanophotonic waveguide. Taking advantage of the inherent link between the local polarization and the propagation direction of the guided light~\cite{petersen2014chiral,mitsch2014quantum}, also referred to as spin-momentum locking~\cite{bliokh2015spin}, $\sigma^-$ ($\sigma^+$) polarized Cs transitions couple to the forward (backward) propagating mode. Depending on the propagation direction of the signal field in the waveguide, this allows us, in conjunction with an external pump field, to address separate $\Lambda$-type atomic energy level schemes in a two-photon Raman configuration. When suitably preparing the atomic spin state, only one of the two $\Lambda$ systems features population inversion, thereby enabling propagation direction-dependent Raman gain. We realize one order of magnitude larger gain for the forward-propagating signal field than for the backward-propagating one. Moreover, we show that the direction in which amplification occurs reverses when flipping the spin of the atoms. Finally, we stabilize the atomic spin by inducing a tensor light shift and demonstrate magnetic field-free operation of the non-reciprocal amplifier. 

Fig.~\ref{fig:setup} shows a schematic of the experimental setup. We work with a pair of counter-propagating fundamental modes of a cylindrical nanophotonic waveguide, which is realized as the waist of a tapered optical fibre. The Cs atoms are held at a distance of ${\sim}\SI{230}{nm}$ from the optical nanofibre surface using a two-colour optical dipole trap and coupled to the nanofibre-guided modes via their evanescent fields~\cite{vetsch2010optical}. We launch the signal field at a wavelength of \SI{852}{\nano\meter} from either one side (port 1) or the other side (port 2) into the tapered fibre. We note, however, that our scheme also works when two fields are launched into port 1 and 2 simultaneously (see Supplementary Section~$1$). We refer to the corresponding power transmission coefficients as $T_{1\rightarrow 2}$ and $T_{2\rightarrow 1}$, respectively. Due to the strong transverse confinement of the signal field in the nanofibre section, a strong component of the electric field along the direction of propagation occurs. This leads to a local polarization at the position of the atoms that is almost perfectly circular and lies in the plane containing the atoms and the nanofibre axis ($x$-$y$-plane) \cite{mitsch2014quantum}. Specifically, the calculated overlap of the signal field at the position of the atoms with $\sigma^-$ ($\sigma^+$) polarization is \SI{92}{\%} (\SI{8}{\%}) when propagating in the $1\rightarrow 2$ direction (quantization axis: $+z$). When propagating in the $2\rightarrow 1$ direction, these specified overlaps of the signal field with $\sigma^-$ and $\sigma^+$ polarization interchange.

\begin{figure}[tb]
	\includegraphics[width=1\linewidth]{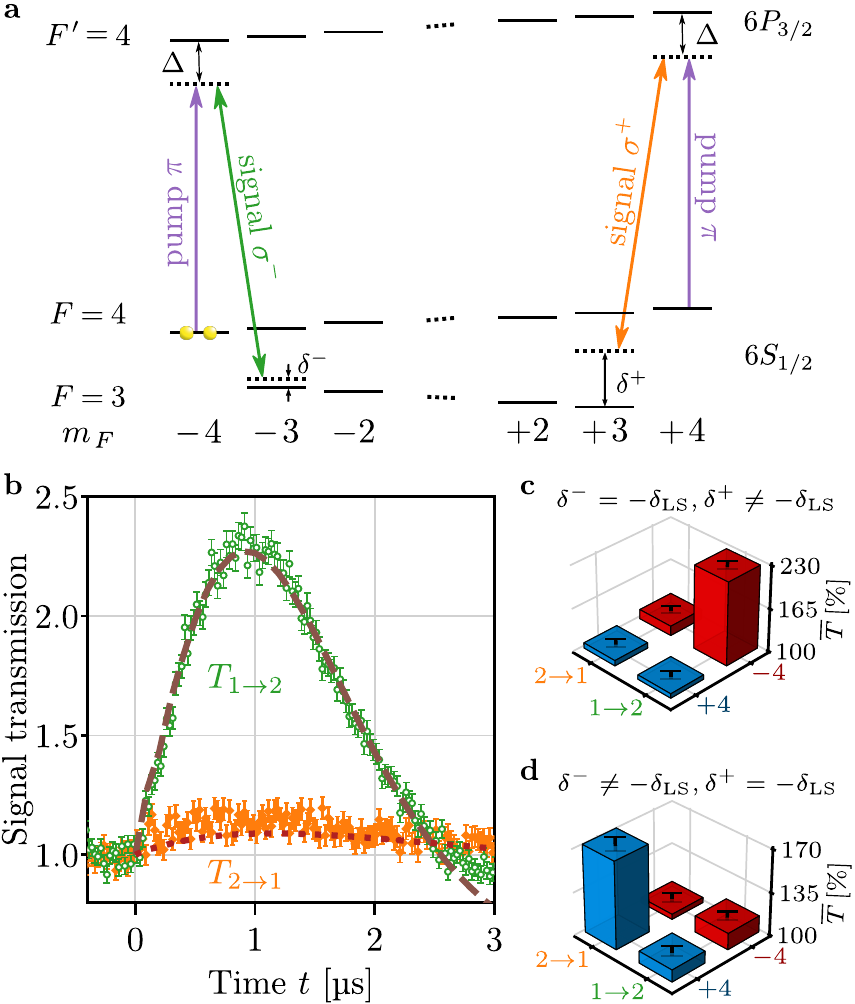}
	\caption{\textbf{Non-reciprocal Raman gain.} \textbf{a},~Relevant Cs energy levels in the presence of an externally applied magnetic field, which is along +z in Fig.~\ref{fig:setup}. The pump field is detuned by $\Delta$ from the $\ket{6S_{1/2}, F=4} \rightarrow \ket{6P_{3/2}, F^\prime=4}$ transition. The $\sigma^\pm$-polarized signal fields couple the levels $\ket{6S_{1/2}, F=3, m_F=\pm3}$ and $\ket{6P_{3/2}, F^\prime=4, m_{F^\prime}=\pm4}$, respectively, and are detuned by $\delta^\pm$ from the two-photon resonance. In combination with their direction-dependent local polarization, this enables non-reciprocal Raman amplification of the signal field. \textbf{b},~Measured transmissions $T_{1\rightarrow 2}$ (green circles) and $T_{2\rightarrow 1}$ (orange diamonds) as a function of time. The atoms are prepared in $\ket{6S_{1/2}, F=4, m_F = -4}$, and the pump field is switched on at $t = \SI{0}{\micro\second}$. We observe that $T_{1\rightarrow 2}$ increases while $T_{2\rightarrow 1}$ remains nearly unchanged, evidencing non-reciprocal gain. The lines are model predictions (see main text). Here and in all following figures, the error bars indicate the $1\sigma$-uncertainty assuming Poissonian statistics for the underlying photon counting. \textbf{c},~Average transmissions, $\overline{T}_{1 \rightarrow 2}$ and $\overline{T}_{2 \rightarrow 1}$, for atoms prepared in $\ket{F=4, m_F = -4}$ or $\ket{F=4, m_F = +4}$ of the $6S_{1/2}$-manifold when the two-photon detuning is set to zero for the $\Lambda^-$ system. \textbf{d},~Same as in \textbf{c} when setting the two-photon detuning for the $\Lambda^+$ system to zero. Non-reciprocal gain can be clearly observed in both cases.
	}
	\label{fig:figure_2_combined}
\end{figure} 

We illuminate the atoms from the side with a $\pi$-polarized free-space pump laser field that propagates in the $+x$-direction and, hence, cannot break reciprocity. In conjunction with the inherent link between the local polarization and the propagation direction of the signal field, this then couples states of different $\Lambda$ configurations, depending on the signal propagation direction (see Fig.~\ref{fig:figure_2_combined}\textbf{a}). For signal propagation from $1\rightarrow2$, the $\Lambda$ configuration comprises a $\pi$ and $\sigma^-$ transition and is denoted $\Lambda^-$. For the other propagation direction, it comprises a $\pi$ and $\sigma^+$ transition and is called $\Lambda^+$. For the following, the relevant $\Lambda^\pm$ systems are formed by the three states \{$\ket{6S_{1/2}, F=4, m_F=\pm4}$, $\ket{6S_{1/2}, F=3, m_F=\pm3}$, $\ket{6P_{3/2},F'=4, m_{F'}=\pm4}$\}, respectively.

When the pump and signal fields fulfill the dressed-atom two-photon resonance, the dynamics that results from these $\Lambda$ couplings only depends on the two-photon Rabi frequency and the decoherence rates. For nanofibre-trapped cold atoms, Raman cooling has been shown recently~\cite{ostfeldt2017dipole}, and the in-trap atomic motion has been measured using Raman coupling~\cite{markussen2020measurement}. Generally, the resonant two-photon Rabi frequency is given by $\Omega_\mathrm{2p}=\Omega_\mathrm{s}\Omega_\mathrm{p}/(2\Delta)$, where $\Omega_\mathrm{s}$ and $\Omega_\mathrm{p}$ are the single-photon Rabi frequencies of the signal and pump field, respectively, and $\Delta$ is the one-photon detuning from the excited state. When the latter can be adiabatically eliminated~\cite{brion2007adiabatic}, and when $\Omega_\mathrm{2p}$ is much larger than the decoherence rates, the dynamics is coherent and two-photon Rabi oscillations of the ground state populations occur. The underlying stimulated Raman scattering leads to an energy transfer from the pump to the signal field while population transfer from $\ket{F=4}$ to $\ket{F=3}$ of the $6S_{1/2}$-manifold takes place. In this case, the signal field experiences gain. Conversely, while population is transferred from $\ket{F=3}$ to $\ket{F=4}$, the signal field experiences loss. In this coherent regime, gain and loss thus vary periodically with time, with a period that is given by $2\pi/\Omega_\mathrm{2p}$. In the incoherent regime, i.e., when $\Omega_\mathrm{2p}$ is much smaller than the decoherence rates, the dynamics can be described by rate equations. There, the gain is proportional to the population difference between $\ket{F=4}$ and $\ket{F=3}$, and the small-signal gain is independent of the signal power, $P_\mathrm{s}$. This is in contrast to the coherent regime, where the maximum gain is inversely proportional to $\Omega_\mathrm{2p}$ and, thus, proportional to $1/\sqrt{P_\mathrm{s}}$.

We now show that the elements introduced above allow non-reciprocal amplification that is controlled by the atomic spin. To this end, we prepare atoms on only one side of the nanofibre (see Methods). This step is performed at an offset magnetic field along $+z$ of ${\sim}\SI{0.5}{G}$ and yields atoms cooled close to the motional ground state, dominantly in the state $\ket{6S_{1/2},F=4, m_F = -4}$~\cite{meng2018near}. The offset magnetic field is then ramped to ${\sim}\SI{7}{G}$ in order to stabilize the $m_F$-state~\cite{dareau2018observation}. We set the one-photon detuning of the pump laser to $\Delta \approx - 2 \pi \times \SI{82}{\mega\hertz}$ and tune the signal field to the light-shifted two-photon resonance, $\delta^-+\delta_\mathrm{LS} = 0$. Here, $\delta_\mathrm{LS} = [(\Omega_\mathrm{p}^2+\Delta^2)^{1/2}-|\Delta|]/2$ is the pump field-induced light shift of the $\ket{6S_{1/2},F=4, m_F=\pm4}$ states and $\Omega_\mathrm{p} \approx 2 \pi \times \SI{20}{\mega\hertz}$ the Rabi frequency of the pump field. We estimate a signal field Rabi frequency of $\Omega_\mathrm{s} \approx 2 \pi \times \SI{1.6}{\mega\hertz}$ for atoms in the minimum of the trapping potential (see Tab.~\ref{tab:theoretical_values} for an overview of relevant parameters).

For reference, we first only switch on the signal field, which does not couple to the atoms in $\ket{6S_{1/2},F=4,m_F=-4}$, and measure the optical power transmitted through the waveguide, yielding ${\sim}\SI{9}{\pico\watt}$. At time $t=\SI{0}{\micro\second}$, we then switch on the pump field and thereby establish the Raman coupling. Figure~\ref{fig:figure_2_combined}\textbf{b} shows the measured signal transmission for propagation from $1 \to 2$ and ${\sim}1400$ trapped atoms (green circles). Initially, it increases linearly and then reaches a maximum of \SI{2.38(6)}{} after $t = \SI{0.9}{\micro\second}$, only slightly less than a quarter of a period of the two-photon Rabi oscillations at a frequency of $\Omega_\mathrm{2p} \approx 2 \pi \times \SI{200}{\kilo\hertz}$. Amplification of the signal field propagating from $1\rightarrow2$ prevails up to $t=\SI{2.5}{\micro\second}$. From $t\approx \SI{2.5}{\micro\second}$ onwards, we observe $T_{1\rightarrow 2}<1$ (see Supplementary Section~$2$). In combination, these two observations evidence that our system operates in a partially coherent regime. For our settings, the main source of decoherence is off-resonant scattering of the pump field which initially occurs at a rate of ${\sim}\SI{500}{\kilo\hertz}$ and then decreases to ${\sim}\SI{100}{\kilo\hertz}$. The modelled evolution of the signal transmission is shown by the dashed line and agrees well with the data. It is obtained by solving the time-dependent Master equation for each atom (see Methods). 

In the other direction, we only observe a small increase of the transmission (orange diamonds and dotted theory line), which reaches a maximum of $T_{2\rightarrow 1}=\SI{1.20(5)}{}$. This residual gain mainly arises from a small $\Lambda^-$ coupling of the $2 \to 1$ signal field due to its ${\sim}8$~\% overlap with $\sigma^-$ polarization. We checked that spontaneous Raman scattering of the pump laser beam into the nanofibre-guided mode contributes at most \SI{11}{\%} of the detected signals. We thus clearly observe non-reciprocal Raman gain.

\begin{figure}[tb]
	\centering
	\includegraphics[width=1\linewidth]{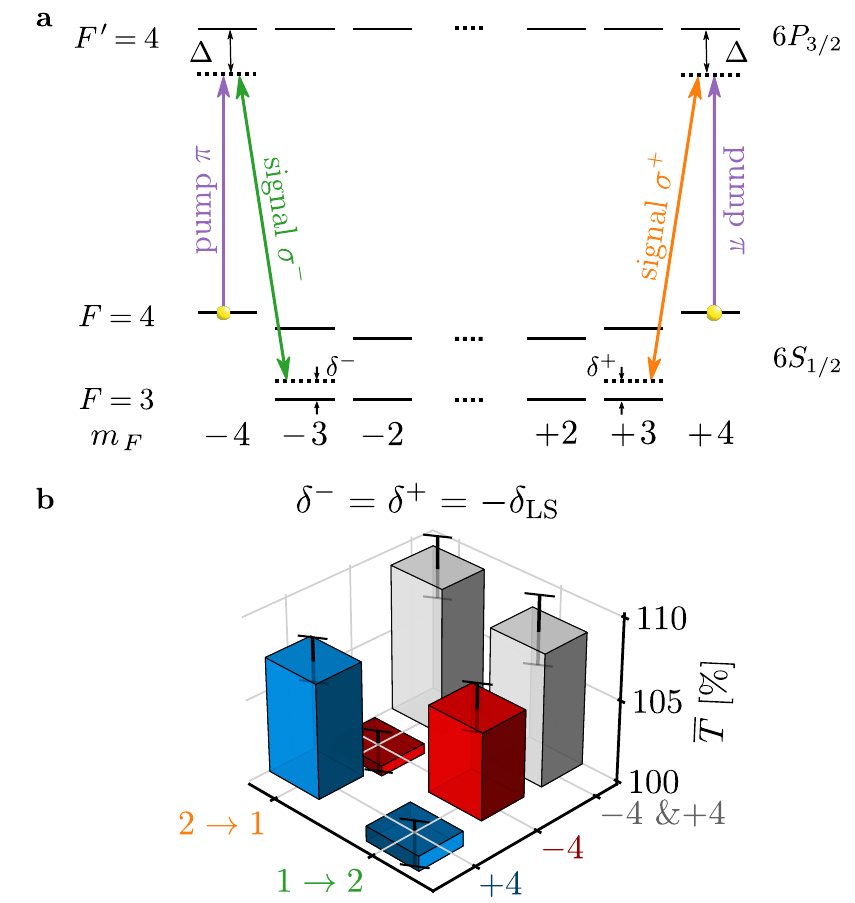}
	\caption{\textbf{Magnetic field-free non-reciprocal Raman gain.} \textbf{a},~We now use a $\pi$-polarized laser field to stabilize the atomic spin state via the tensor light shift. For this configuration, the two-photon detunings, $\delta^+$ and $\delta^-$, coincide (see main text). \textbf{b},~Setting the gain direction of the magnetic field-free amplifier: At two-photon resonance, amplification occurs in both directions when the atoms are prepared in a statistical mixture of $\ket{F=4, m_F = -4}$ and $\ket{F=4, m_F = +4}$ of the $6S_{1/2}$ manifold (grey bars). For atoms prepared in $\ket{F=4, m_F = -4}$ only, we observe gain exclusively for the $1 \rightarrow 2$ direction. When the atoms are prepared in $\ket{F=4, m_F = +4}$, amplification occurs in the $2\to1$ direction.}
	\label{fig:tensor_amplification}
\end{figure}

In Fig.~\ref{fig:figure_2_combined}\textbf{c}, we plot the mean signal transmissions, $\overline{T}_{1 \rightarrow 2}$ and $\overline{T}_{2 \rightarrow 1}$, averaged from $t=\SIrange{0.7}{1.2}{\micro\second}$. The red bars are derived from the data shown in Fig.~\ref{fig:figure_2_combined}\textbf{b}, where the atoms are prepared in the $\ket{F=4, m_F=-4}$ state of the $6S_{1/2}$ manifold. In order to reverse the direction of the Raman gain, we now switch from $\Lambda^-$ to $\Lambda^+$ coupling by preparing the atoms in $\ket{F=4, m_F=+4}$ (see Methods). However, when only switching the initial atomic spin state, no gain is observed in either direction (blue bars in Fig.~\ref{fig:figure_2_combined}\textbf{c}). This is because of the Zeeman shifts of the energy levels of the $6S_{1/2}$ manifold, which leads to a large two-photon detuning, $\delta^+$, for the $\Lambda^+$ system (see Fig.~\ref{fig:figure_2_combined}\textbf{a}). By adjusting the signal frequency to compensate for this Zeeman shift, we restore the two-photon resonance for the $\Lambda^+$ system and, indeed, observe non-reciprocal Raman gain in the opposite ($2\to1$) direction, (blue bars in Fig.~\ref{fig:figure_2_combined}\textbf{d}). For this setting of the signal detuning, in turn, no gain is observed for either direction when atoms are prepared in $\ket{F=4, m_F=-4}$ (red bars).

We now demonstrate that non-reciprocal Raman gain can also be achieved in our system without an applied magnetic field. However, in this case, we have to stabilize the $m_F$ states by means of another mechanism against depopulation and dephasing due to spin-motion coupling~\cite{dareau2018observation} and stray magnetic fields. To this end, we employ an additional $\pi$-polarized laser field. This beam co-propagates with the pump beam, is about $2 \pi \times \SI{90}{\mega\hertz}$ red-detuned to the $\ket{6S_{1/2},F=4} \rightarrow \ket{6P_{1/2},F^\prime=3}$ transition at a wavelength of \SI{894}{\nano\meter}, and has an intensity of \SI{280}{\milli\watt\per\centi\meter\squared}. It induces a tensor light shift (TLS) for the $F=4$ manifold of the $6S_{1/2}$ ground state while its effect on the $F=3$ ground state manifold is negligible (see Fig.~\ref{fig:tensor_amplification}\textbf{a}). For our experimental parameters, we estimate the differential TLS between the $\ket{F=4, m_F = \pm 4}$ and the $\ket{F=4, m_F = \pm 3}$ state to be about $2 \pi \times \SI{2}{\mega\hertz}$, large enough to stabilize the $\ket{F=4, m_F = \pm 4}$ states. Crucially, the TLS is proportional to $m_F^2$, so that the two-photon resonances in the $\Lambda^+$ and $\Lambda^-$ configurations now coincide, $\delta^+=\delta^-$ (see Supplementary Section~$3$ for experimental confirmation). 

We first demonstrate reciprocal gain in this setting. For this, we prepare atoms in a statistical mixture of $\ket{6S_{1/2},F=4, m_F = +4}$ and $\ket{6S_{1/2},F=4, m_F = -4}$ (see Methods). The mean transmissions, $\overline{T}_{1 \rightarrow 2}$ and $\overline{T}_{2 \rightarrow 1}$, averaged from $t=\SIrange{0.15}{0.9}{\micro\second}$ are shown in Fig.~\ref{fig:tensor_amplification}\textbf{b} (grey bars). They coincide within the error bars, confirming reciprocal amplification. In these measurements, the average gain is smaller than ${\sim}10$\%, limited by a less effective state preparation and roughly five times larger decoherence rates, e.g., due to scattering of the imperfectly $\pi$-polarized TLS field. In order to obtain non-reciprocal gain, we now prepare atoms selectively in one of the two outermost Zeeman states, either $\ket{F=4, m_F =-4}$ or $\ket{F=4, m_F=+4}$. For $\ket{F=4, m_F=-4}$, we only observe gain in the $1 \to 2$ direction (red bars). For $\ket{F=4,m_F =+4}$, gain only occurs for the $2 \to 1$ direction (blue bars). This confirms that the atomic spin rather than an applied magnetic field breaks reciprocity and allows us to control the direction of amplification in our experiment.

While our experiment operated in a pulsed regime, the mechanism also allows for continuous-wave gain by means of a suitable repump laser field. Moreover, our results enable the investigation of cold-atom lasing~\cite{guerin2008mechanisms} with spin-controlled directionality of the atom--light coupling. Given the high level of control available with our experimental platform, it lends itself to studying quantum thermodynamics in the presence of non-reciprocal interactions~\cite{braak2020fermi, loos2020irreversibility}. Finally, our amplification scheme could also be implemented with other emitters with a suitable level scheme coupled to spin-momentum locked waveguides~\cite{lodahl2017chiral}, e.g., in circuit quantum electrodynamics, where non-reciprocal magnetic field-free amplification at the quantum level is a highly sought capability \cite{clerk2010introduction, gu2017microwave}.

\bibliography{bibliography}
 

\section*{Methods}

\textbf{Nanofibre-based dipole trap.}
We trap laser-cooled Cs atoms in a nanofibre-based two-colour optical dipole trap~\cite{vetsch2010optical}. The nanofibre is implemented as the waist of a tapered step-index optical fibre. The nanofibre part is ${\sim}\SI{5}{\milli\meter}$ long and has a diameter of ${\sim}\SI{500}{\nano\meter}$. The blue-detuned trapping light field has a free-space wavelength of $\lambda=\SI{760}{\nano\meter}$, a power of ${\sim}\SI{20.5}{\milli\watt}$, and is launched as a running wave into the fibre. The red-detuned standing-wave field with $\lambda=\SI{1064}{\nano\meter}$ and a total power of ${\sim}\SI{2.4}{\milli\watt}$ is also guided in the nanofibre. All trapping light fields propagate as $\mathrm{HE}_{11}$ modes and are quasilinearly polarized~\cite{lekien2004field}. The blue-detuned trapping light field's polarization plane is orthogonal to the polarization plane of the red-detuned light field. As a result of these settings, two diametric arrays of optical trapping sites form along the nanofibre. These dipole trap minima are located ${\sim}\SI{230}{\nano\meter}$ away from the nanofibre surface.

\textbf{Preparation of the atomic ensemble.} 
The atoms are loaded from a magneto-optical trap into the nanofibre-based trap via an optical molasses stage~\cite{vetsch2010optical}. The collisional blockade effect limits the maximum number of atoms per trapping site to at most one \cite{schlosser2002collisional}. The filling of the trapping sites is random, with an average filling factor of ${\sim}\SI{20}{\%}$~\cite{vetsch2012nanofiber}. In order to only work with atoms on one side of the fibre, we first have to remove the atoms stored in one of the two diametric arrays. We achieve this by side-selective degenerate Raman heating~\cite{meng2018near} for \SI{40}{\milli\second}, after which the heated atoms leave the trap. Simultaneously, the atoms on the other side of the fibre are subject to degenerate Raman cooling (DRC) and thus stay in the trap. The cooling turns out to be most efficient for an offset magnetic field of ${\sim}\SI{0.5}{\gauss}$ applied along $+z$ in Fig.~\ref{fig:setup}. In order to infer the number of remaining trapped atoms, we measure the transmission of a fibre-guided laser field while scanning its frequency across the resonance of the Cs $\ket{6S_{1/2}, F=4} \rightarrow \ket{6P_{3/2}, F^\prime=5}$ cycling transition. From a fit of the transmission spectrum, we obtain the optical density (OD) of the atomic ensemble and convert it to the number of trapped atoms, $N$. For the conversion, we independently calibrate the OD per atom using a saturation measurement~\cite{vetsch2010optical}. 

Following this transmission measurement, another sequence of DRC is applied. It is intrinsic to this cooling scheme that the atoms are optically pumped to the $\ket{6S_{1/2}, F=4, m_F=-4}$ state. In order to demonstrate non-reciprocal amplification (cf.~Fig.~\ref{fig:figure_2_combined}) the magnetic field is first ramped from ${\sim}\SI{0.5}{\gauss}$ to ${\sim}\SI{7}{\gauss}$ within \SI{7}{\milli\second}. This guiding offset field ensures that the atoms remain in $\ket{F=4, m_F=-4}$. From there, if the atoms are to be prepared in $\ket{F=4, m_F=+4}$, we perform optical pumping by means of a $\sigma^+$-polarized external laser beam that propagates in the $+z$-direction and couples to the Cs $\ket{6S_{1/2}, F=4} \to \ket{6P_{1/2}, F^\prime=4}$ transition. 

For the experiment on non-magnetic non-reciprocal amplification
(cf.~Fig.~\ref{fig:tensor_amplification}), the state preparation is performed differently. There, we ramp the magnetic field to zero after the DRC stage. Residual magnetic stray fields as well as spin-motion coupling~\cite{dareau2018observation} then distribute the atomic population over all Zeeman states of the $F=4$ ground-state manifold. We then switch on the $\pi$-polarized tensor light shift laser beam which is $2 \pi \times \SI{90}{\mega\hertz}$ red-detuned from the Cs $\ket{6S_{1/2}, F=4} \to \ket{6P_{1/2}, F^\prime=3}$ transition. In addition to inducing level shifts, it optically pumps the atoms into an incoherent mixture of the $\ket{6S_{1/2},F=4, m_F = -4}$ and $\ket{6S_{1/2},F=4, m_F = +4}$ states within \SI{100}{\micro\second}. These two outermost Zeeman states are dark states for the TLS laser field. In order to exclusively prepare atoms in either $\ket{F=4, m_F = -4}$ or $\ket{F=4, m_F = +4}$, we apply a $\sigma^\pm$-polarized optical pumping light field simultaneously with the TLS beam. In this case, the dark-state condition for atoms in $m_F = \mp4$ is lifted, they experience strong recoil heating and are removed from the trap while atoms in $\ket{F=4, m_F = \pm 4}$ are nearly unaffected.

\textbf{Measurement of signal gain and loss.} 
For measuring the signal gain, we probe the atoms in an alternating fashion, i.e.~switching the signal's propagation direction in every run while measuring the transmitted optical power using single-photon counting modules (SPCMs). This reduces the influence of possible long-term drifts of the setup on, e.g., the observed gain asymmetry. For reference, in every experimental run, we also record the photons counts for a guided signal beam that does not couple to atoms in the $\ket{6S_{1/2},F=4}$ ground state for a duration of $\SI{100}{\micro\second}$. In addition, we record the background count rate in the absence of signal and pump laser fields for $\SI{100}{\micro\second}$. In total, one experimental cycle takes ${\sim}\SI{3}{\second}$. We repeat the experiment until a good counting statistics is reached, typically involving $>1000$ cycles. For details on complementary measurements with the signal field sent in both directions through the fibre simultaneously, see Supplementary Section~$1$. 

\textbf{Detection setup and data analysis.} 
On each end of the tapered optical fibre, we employ two identical bandpass filtering systems with a centre wavelength of \SI{\sim 852}{\nano\meter} in order to suppress the trapping fields in the transmitted signal beam. Each system consists of a dichroic mirror and a volume Bragg grating. The filtered signal field hits a 50:50 beam splitter. To suppress optical cross-talk, each beam then passes an additional bandpass filter and is guided to a SPCM via a multimode fibre. The count rate on each detector always stays below \SI{2}{\mega\hertz}, minimizing SPCM saturation. The SPCM detection events are time-tagged and then binned. From this raw data, we infer the signal transmission by correcting both the signal and the reference data for the background and then computing their ratio. The corresponding histogram is shown in Fig.~\ref{fig:figure_2_combined}\textbf{b} and displays the dynamics of the signal transmission.

\begin{table*}[tb]
\caption{\textbf{Overview of system parameters.} We present the parameters used in our model for calculating the signal transmission in the $1 \rightarrow 2$ direction (see dashed line in Fig.~\ref{fig:figure_2_combined}\textbf{b}). For comparison, we provide independently inferred values of those parameters and how those have been obtained.}
\label{tab:theoretical_values}
\begin{tabular}{lllll}
\hline
Parameter & Description & Model parameters & Independent estimate & Inferred from \\ \hline
$\Omega_\mathrm{s}$ & Signal field Rabi frequency & $2 \pi \times \SI{0.95}{\mega\hertz}$ & $2 \pi \times \SI{1.6}{\mega\hertz}$ & \begin{tabular}{@{}l@{}} Measured transmitted signal power and \\ atom--fibre distance \end{tabular} \\ [0.3cm]
$\Omega_\mathrm{p}$ & Pump field Rabi frequency & $2 \pi \times \SI{20.7}{\mega\hertz}$ & $2 \pi \times \SI{19.96(9)}{\mega\hertz}$ & Measured Autler-Townes splitting \\ [0.1cm]
$\Delta$ & One-photon detuning & $2 \pi \times \SI{82}{\mega\hertz}$ & $2 \pi \times \SI{82.3(5)}{\mega\hertz}$ & Measured signal transmission spectrum \\ [0.1cm]
$\delta^-+\delta_\mathrm{LS}$ & \begin{tabular}{@{}l@{}} Detuning from dressed state \\ resonance \end{tabular}  & $2 \pi \times \SI{0}{\mega\hertz}$ & $2 \pi \times \SI{0.00(2)}{\mega\hertz}$ & See Supplementary Section~$3$ \\ [0.3cm]
$N$ & Number of atoms & $1420$ & $1412(168)$ & Measured signal transmission spectrum \\ [0.1cm]
$\mathrm{OD}_{0}$ & \begin{tabular}{@{}l@{}} OD per atom on $a \to e$ \\ transition for $\Omega_\mathrm{p}=0$   \end{tabular} & $0.0131$ & $0.0122(15)$ & Saturation measurement \cite{vetsch2010optical} \\ [0.3cm]
$\gamma_\mathrm{ba}$ & Ground state decoherence rate & $2 \pi \times \SI{0.29}{\mega\hertz}$ & $2 \pi \times \SI{0.47(3)}{\mega\hertz}$ & See Supplementary Section~$3$ \\ [0.1cm]
$\gamma_\mathrm{e}$ & Excited state decay rate & $2 \pi \times \SI{5.225}{\mega\hertz}$ & $2 \pi \times \SI{5.225(8)}{\mega\hertz}$ & Literature value~\cite{patterson2015lifetime} \\ \hline
\end{tabular}
\end{table*}

\textbf{Theoretical description.}
In our theoretical model, we propagate the signal field through an array of atoms which feature three energy levels in a $\Lambda$ configuration and which are driven by the pump field. The atomic ground states $\ket{\mathrm{a}}$ and $\ket{\mathrm{b}}$, and the excited state $\ket{\mathrm{e}}$ are associated with the states $\ket{6S_{1/2}, F=3, m_F=-3}$, $\ket{6S_{1/2}, F=4, m_F=-4}$, and $\ket{6P_{3/2}, F^{\prime}=4, m_{F^{\prime}}=-4}$ of atomic Cs, respectively (cf.~Fig.~\ref{fig:figure_2_combined}a). We model the signal transmission past each atom as $T = |h|^2$ with the transfer function~\cite{sayrin2015storage, fleischhauer2005electromagnetically}
\begin{equation}
h = \exp \left[\mathrm{i} \frac{\mathrm{OD}_{0}}{2} \tilde{\chi}_{\mathrm{ae}}\right],
\label{eq:transmission}
\end{equation}
where $\mathrm{OD}_{0}$ is the resonant single-atom optical depth of the signal field-driven $\ket{\mathrm{a}} \rightarrow \ket{\mathrm{e}}$ transition in absence of a pump field and
\begin{equation}
\tilde{\chi}_{\mathrm{ae}}(t) = \frac{\gamma_\mathrm{e}}{2 \Omega_{\mathrm{s}}} \rho_{\mathrm{ae}}(t, \Omega_{\mathrm{s}})~.
\label{eq:chi_rho}
\end{equation}
Here, $\gamma_\mathrm{e}$ is the decay rate of the excited state population, and  $\rho_{\mathrm{ae}}(t, r)$ is the time-dependent density matrix element of the atomic density operator, $\hat \rho$. The dynamics of $\rho_{\mathrm{ae}}$ is governed by the Lindblad master equation for $\hat \rho$, which we solve numerically using QuTiP~\cite{johansson2013qutip}.

We then numerically propagate the signal field through the ensemble. The signal field amplitude launched into the fibre is assumed to be constant in time, thus driving the first atom with a fixed Rabi frequency. The second atom is then driven by the superposition of the initial signal field and the modulated field radiated by the first atom. More quantitatively, the Rabi frequency of the signal light field at the second atom is given by $\Omega_{\mathrm{s, 2}}(t)=\Omega_{\mathrm{s,1}} h_1(t)$. Here, $\Omega_{\mathrm{s,1}}$ is the Rabi frequency of the signal light at the first atom. The third atom is then driven by the superposition of the initial field and the modulated fields radiated by the first two atoms, and so on. The dashed line in Fig.~\ref{fig:figure_2_combined}\textbf{b} is the result of this calculation for the set of parameter values listed in Tab.~\ref{tab:theoretical_values}, which are chosen to obtain good agreement with the experimental data. For comparison, Tab.~\ref{tab:theoretical_values} also lists independently determined values of the same parameters, which are in reasonable agreement.

\section*{Acknowledgments}

We thank A.~Husakou, F.~Tebbenjohanns, and J.~Volz  for stimulating discussions and helpful comments. We acknowledge funding by the Alexander von Humboldt Foundation in the framework of the Alexander von Humboldt Professorship endowed by the Federal Ministry of Education and Research. Moreover, financial support from the European Union's Horizon 2020 research and innovation program under grant agreement No.~800942 (ErBeStA) is gratefully acknowledged.

\section*{Author contributions}
C.L.~and S.P.~contributed equally to this work. C.L.~and S.P.~adapted and extended the setup for this experiment and carried out the measurement. S.J.~assisted in the early stages of the experiments. The data was analysed and modelled by C.L., S.P., and P.S. The experiment was conceived and supervised by A.R.~and P.S. All authors contributed to the writing of the manuscript.

\section*{Data availability}
The data that support the plots of this paper as well as the other findings of our study are available from the corresponding author upon reasonable request.

\end{document}



\title{Supplementary information: Atomic spin-controlled non-reciprocal Raman amplification of fibre-guided light}

\author{Sebastian Pucher}\thanks{C.L. and S.P. contributed equally to this work.}

\author{Christian Liedl}\thanks{C.L. and S.P. contributed equally to this work.}

\author{Shuwei Jin}

\author{Arno Rauschenbeutel}

\author{Philipp Schneeweiss}
\email{philipp.schneeweiss@hu-berlin.de}

\affiliation{
 Department of Physics, Humboldt-Universit\"at zu Berlin, 10099 Berlin, Germany
}

\date{\today}

\maketitle

\section{Simultaneous probing with signal light from both directions}
\label{sec:Simultaneous probing}
We now explore whether non-reciprocal amplification prevails even when sending signal fields simultaneously into both ports of the waveguide. In all other aspects, we follow the same procedure as for the experiment shown in Fig.~\textbf{2}b of the main manuscript, where we prepare the atoms in the $\ket{F=4, m_F=-4}$ ground state. 

\begin{figure}[h]
	\includegraphics[width=0.4\columnwidth]{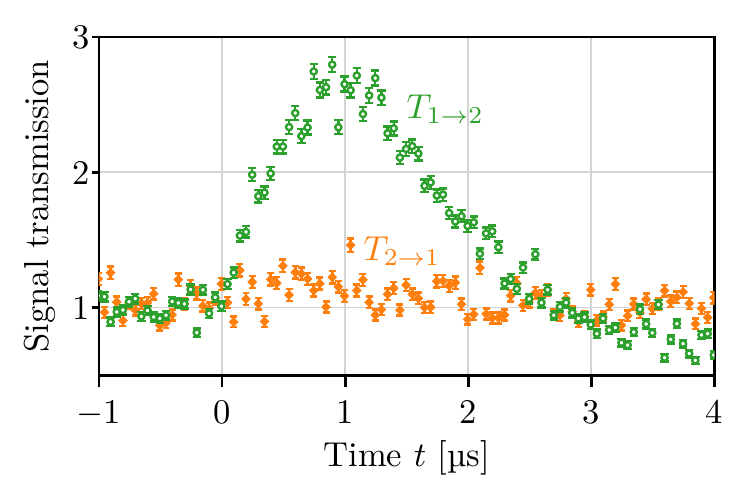}
	\caption{\textbf{Non-reciprocal Raman gain when simultaneously probing from both directions.} Measured signal transmission coefficients $T_{1\rightarrow2}$ (green circles) and $T_{2\rightarrow1}$ (orange diamonds) as a function of time. Here, we launch the signal fields into both ports of the waveguide simultaneously. We clearly observe non-reciprocal amplification also under these conditions.
}
	\label{fig:simultaneous_probing}
\end{figure}

In Fig.~\ref{fig:simultaneous_probing}, we present the resulting signal transmissions as a function of time. We observe a non-reciprocal amplification, similar to the experiments presented in the main manuscript. The signal transmission reaches a maximum value of ${\sim}2.6$ in the $1\rightarrow2$ direction and is almost unchanged in the $2\rightarrow1$ direction. This confirms that our scheme for non-reciprocal amplification also works when sending the signal field into both directions at the same time.

\section{Extended Fig.~2a: Signal transmission dynamics}
\begin{figure}[h]
	\includegraphics[width=1.0\columnwidth]{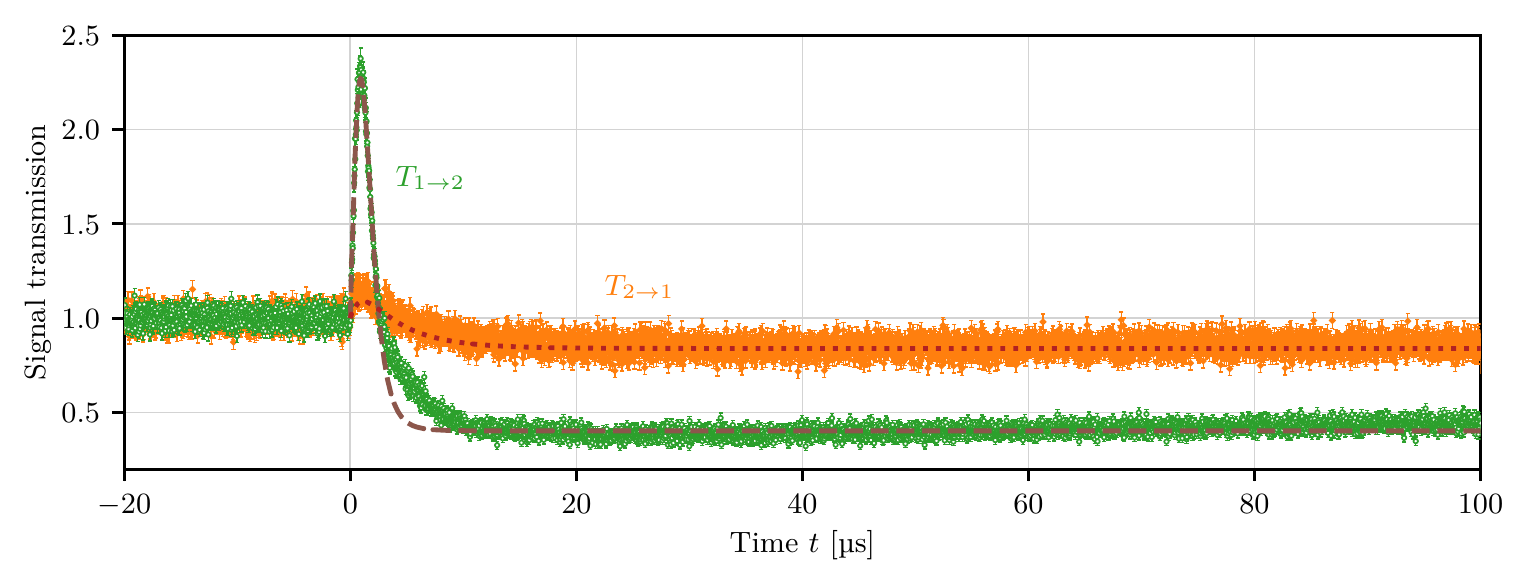}
	\caption{\textbf{Extended Fig.~2a of the main manuscript.} Measured signal transmissions $T_{1\rightarrow 2}$ (green circles) and $T_{2\rightarrow 1}$ (orange diamonds) as a function of time. The atoms are prepared in $\ket{F=4, m_F = -4}$. Before switching on the pump field, we measure constant transmitted signal powers in both directions. After switching on the pump field at $t = \SI{0}{\micro\second}$, $T_{1\rightarrow 2}$ increases while $T_{2\rightarrow 1}$ remains nearly unchanged. At $t \approx \SI{2.5}{\micro\second}$, the signal gain turns into loss, which persists until the atoms are pumped to uncoupled states or heated out of the trap at even longer times (not shown). The lines are model predictions. The error bars indicate the $1\sigma$-uncertainty assuming Poissonian statistics for the underlying photon counting.}
	\label{fig:time_evolution_broad}
\end{figure}

Here, we discuss the dynamics of the signal transmission in more detail. In Fig.~\ref{fig:time_evolution_broad}, we present the measured signal transmissions $T_{1\rightarrow 2}$ and $T_{2\rightarrow 1}$ of Fig.~2\textbf{b} of the main manuscript including data for longer times. Before switching on the pump laser ($t < \SI{0}{\micro\second}$ in Fig.~\ref{fig:time_evolution_broad}), we observe constant transmitted signal powers, which serve as a reference, i.e.~we set $T_{1\rightarrow 2}=T_{2\rightarrow 1}=1$ for $t < \SI{0}{\micro\second}$. At $t = \SI{0}{\micro\second}$, we switch on the pump field. As discussed in the main manuscript, we then observe non-reciprocal amplification up to $t \approx \SI{2.5}{\micro\second}$. Subsequently, the signal field experiences loss due to two-photon absorption, resulting in a transmission $<1$ in both directions. After $t \approx \SI{10}{\micro\second}$, we reach a quasi-steady state.

We manually fit our theory prediction to the experimentally obtained transmissions. The match between theory and experiment is less good for the larger time interval presented in Fig.~\ref{fig:time_evolution_broad} compared to the time interval shown in Fig.~2\textbf{b} of the main manuscript. The deviation might arise from light-induced dipole forces exerted by the pump laser field onto the nanofibre-trapped atoms. Such forces are not taken into account in our model and may excite a centre-of-mass oscillation of the atomic ensemble with a period of ${\sim}\SI{6}{\micro\second}$, estimated from the inverse of the calculated radial trap frequency, $\nu_r= 2 \pi \times \SI{160}{\kilo\hertz}$. Furthermore, our theory neglects loss of population out of the $\Lambda$-system due to excitation and decay to other Cs levels. Finally, we neglect the state dependence of the trapping potential~\cite{lekien2013state}, which can, e.g., lead to excess heating and a concurrent change of the coupling strength of the atom to the nanofibre-guided mode.

\section{Frequency dependence of signal transmission}
\begin{figure}[h]
	\includegraphics[width=1\columnwidth]{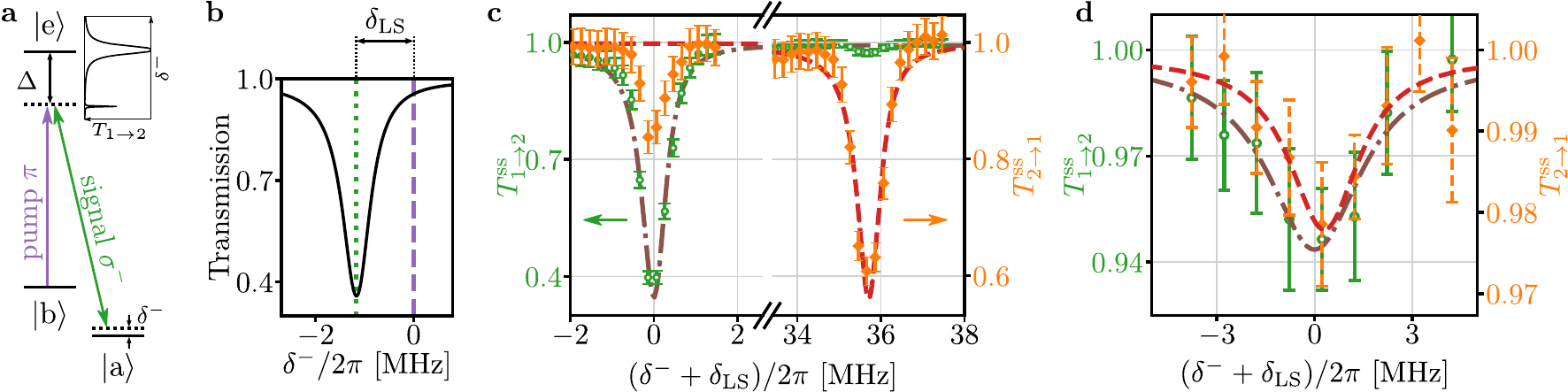}
	\caption{\textbf{Level scheme and signal transmission spectra.} \textbf{a}, Levels and transitions of our $\Lambda$-system. The inset shows schematically the theoretically expected steady-state signal transmission when we scan $\delta^-$ over the one-photon and two-photon resonances. \textbf{b}, Zoom of the calculated signal transmission spectrum. Due to the light shift induced by the pump field, the dressed-atom two-photon resonance (green dotted line) is shifted by $\delta_\mathrm{LS}$ with respect to the bare-atom two-photon resonance (purple dashed line). \textbf{c}, Measured quasi-steady-state signal transmissions, averaged between $t=\SI{20}{\micro\second}$ and $t=\SI{50}{\micro\second}$, as a function of the two-photon detuning. When the initial atomic state is $\ket{F=4, m_F = -4}$ ($\ket{F=4, m_F = +4}$), we observe loss in $T^{\mathrm{ss}}_{1\rightarrow 2}$ ($T^{\mathrm{ss}}_{2\rightarrow 1}$), see green circles (orange diamonds). The two-photon resonance condition is met at different values of $\delta^-$, due to the presence of a finite offset magnetic field. We use our model to perform fits (see text). The fits of the frequency dependence of $T^{\mathrm{ss}}_{1\rightarrow 2}$ (brown dash-dotted line) as well as that of $T^{\mathrm{ss}}_{2\rightarrow 1}$ (red dashed line) agree well with the experimental data. \textbf{d} Measured quasi-steady-state signal transmission in the non-magnetic scheme. For both propagation directions of the signal field, we observe the maximum of the two-photon loss at the same value of $\delta^{-}$. Fits unveil the resonances at $\delta^- + \delta_{\mathrm{LS}} = 2 \pi \times \SI{0.0(2)}{\mega\hertz}$ for $T^{\mathrm{ss}}_{1 \rightarrow 2}$ and $\delta^- + \delta_{\mathrm{LS}} = 2 \pi \times \SI{0.3(2)}{\mega\hertz}$ for $T^{\mathrm{ss}}_{2 \rightarrow 1}$. Clearly, the resonances are separated by much less than their respective widths. From the separation of the resonance positions, we extract a residual magnetic field of \SI{0.06(6)}{\gauss}.}
	\label{fig:freq_scan}
\end{figure}

To study the frequency dependence of the signal gain and loss, we scan the two-photon detuning $\delta^-$ and measure $T_{1\rightarrow 2}$. As illustrated in Fig.~\ref{fig:freq_scan}\textbf{a}, $\delta^-$ is defined with respect to the bare states of the atom. However, the pump field induces a light shift of the $\ket{b} \to \ket{e}$ transition. This leads to a shift of the resonance condition at which two-photon Raman gain and loss occur. This shift is obtained from the dressed-state eigenenergies for the pump-coupled $\ket{b} \to \ket{e}$ transition and given by
\begin{equation}
\delta_\mathrm{LS} = \frac{1}{2}\left(\sqrt{\Omega_\mathrm{p}^2+\Delta^2}-|\Delta|\right)~,
\end{equation}
where $\Delta \approx -2 \pi \times \SI{82}{\mega\hertz}$ is the one-photon detuning of the pump laser from the excited state, and $\Omega_\mathrm{p} \approx 2 \pi \times \SI{20}{\mega\hertz}$ is the pump field Rabi frequency. Figure~\ref{fig:freq_scan}\textbf{b} illustrates this shift of the resonance.

\noindent We model the signal transmission as $T = |h|^2$, where the transfer function $h$ is given by~\cite{sayrin2015storage}
\begin{equation}
h =  \exp \left[\mathrm{i} \frac{\mathrm{OD}}{2} \tilde{\chi}_{\mathrm{ae}}\right]~.
\label{eq:transfer_function}
\end{equation}
Here, $\mathrm{OD}$ is the resonant optical depth on the $\ket{a} \to \ket{e}$ transition in the absence of a pump field and, under steady-state conditions, $\tilde{\chi}_{\mathrm{ae}}$ is given by~\cite{fleischhauer2005electromagnetically, sayrin2015storage}
\begin{equation}
\tilde{\chi}_{\mathrm{ae}} = \frac{\gamma_{\mathrm{e}}}{2}
\times\left[\frac{4 \delta^{-} \left(\left|\Omega_{\mathrm{p}}\right|^{2}-4 \delta^{-} \Delta\right)-4 \Delta \gamma_{\mathrm{ab}}^{2}}{|\left|\Omega_{\mathrm{p}}\right|^{2}+\left.\left(\gamma_{\mathrm{e}}+ \mathrm{i} 2 \Delta\right)\left(\gamma_{\mathrm{ab}}+ \mathrm{i} 2 \delta^{-}\right)\right|^{2}}\right.
\left.
+ \mathrm{i} \frac{8 (\delta^{-})^{2} \gamma_{\mathrm{e}}+2 \gamma_{ab}\left(\left|\Omega_{\mathrm{p}}\right|^{2}+\gamma_{\mathrm{ab}} \gamma_{\mathrm{e}}\right)}{|\left|\Omega_{\mathrm{p}}\right|^{2}+\left.\left(\gamma_{\mathrm{e}}+ \mathrm{i} 2 \Delta\right)\left(\gamma_{\mathrm{ab}}+ \mathrm{i} 2 \delta^{-} \right)\right|^{2}}\right]~,
\label{eq:Fleischhauer}
\end{equation}
where $\gamma_\mathrm{e}$ is the excited state decay rate and $\gamma_{\mathrm{ab}}$ is the ground state decoherence rate.

In Fig.~\ref{fig:freq_scan}\textbf{c}, we plot the quasi-steady-state signal beam transmission, $T^{\mathrm{ss}}_{1 \rightarrow 2}$, measured between $t=\SI{20}{\micro\second}$ and $t=\SI{50}{\micro\second}$ (green circles). This corresponds to a time interval during which the signal field experiences two-photon loss. We see a clear dip in the transmission spectrum, indicating the two-photon resonance of the atoms that are dressed by the pump laser field. We use the above equations to perform a fit of the $T^{\mathrm{ss}}_{1 \rightarrow 2}$ data. Our free fit parameters are $\gamma_{\mathrm{ab}}$, $\mathrm{OD}$, and $\Delta$. We find an excellent agreement between the fit and the measured transmission spectrum, yielding a ground state decoherence rate of $\gamma_\mathrm{ab} = 2 \pi \times \SI{0.47(3)}{\mega\hertz}$. Moreover, we derive a detuning of $\delta^- + \delta_{\mathrm{LS}} = 2 \pi \times \SI{0.00(1)}{\mega\hertz}$ from the values of the fit parameters. In order to change the direction in which gain and loss occur, we now prepare the atoms in state $\ket{F=4, m_F = +4}$ via optical pumping. For this setting, we obtain the two-photon loss spectrum from a measurement of the quasi-steady-state signal transmission $T^{\mathrm{ss}}_{2 \rightarrow 1}$ as a function of $\delta^- + \delta_{\mathrm{LS}}$ (orange diamonds in Fig.~\ref{fig:freq_scan}\textbf{c}). A peak with a width corresponding to a ground state decoherence rate of $\gamma_\mathrm{ab} = 2 \pi \times \SI{0.51(8)}{\mega\hertz}$ is apparent at $\delta^- + \delta_{\mathrm{LS}} = 2 \pi \times \SI{35.71(3)}{\mega\hertz}$. From the frequency difference of the dressed state resonances, we infer a magnetic field strength at the position of the atoms of \SI{7.29(1)}{\gauss}. This agrees with the independently calibrated homogeneous offset magnetic field.

We now carry out measurements without offset magnetic field while stabilizing the spin-polarization of the atoms using tensor-light shifts. In order to determine the two-photon resonance, we measure $T^{\mathrm{ss}}_{1 \rightarrow 2}$ and $T^{\mathrm{ss}}_{2 \rightarrow 1}$ as a function of $\delta^-$ (see Fig.~\ref{fig:freq_scan}\textbf{d}). Fits unveil the light shifted two-photon resonances at $\delta^- + \delta_{\mathrm{LS}} = 2 \pi \times \SI{0.0(2)}{\mega\hertz}$ and $\delta^- + \delta_{\mathrm{LS}} = 2 \pi \times \SI{0.3(2)}{\mega\hertz}$ for $T_{1 \rightarrow 2}$ and $T_{2 \rightarrow 1}$, respectively. Thus, the resonances coincide within their linewidths ($1 \rightarrow 2$ direction: $\gamma_\mathrm{ab} = 2 \pi \times \SI{3.4(5)}{\mega\hertz}$; $2 \rightarrow 1$ direction: $\gamma_\mathrm{ab} = 2 \pi \times \SI{2.6(6)}{\mega\hertz}$). From the observed shift of the resonances, we extract a residual magnetic field of \SI{0.06(6)}{\gauss}. The ground state decoherence rates observed for the magnetic-field free scheme are larger than those obtained in the measurements with offset magnetic fields. This is possibly due to the inhomogeneous intensity distribution and the imperfect $\pi$-polarization of the light field that induces the tensor light shifts.

\bibliography{bibliography}